\documentclass[10pt,conference,USletter]{IEEEtran}
\UseRawInputEncoding

\usepackage{cite}
\usepackage{amsmath,amssymb,amsfonts}

\usepackage{graphicx}
\usepackage{textcomp}
\usepackage{xcolor}

\usepackage{algorithm}
\usepackage{algpseudocode}

\usepackage{float}
\usepackage{caption}
\usepackage{subcaption}
\usepackage{url}
\usepackage{paralist}

\renewcommand{\baselinestretch}{0.96}

\usepackage{soul}

\usepackage{tikz}
\usetikzlibrary{positioning, calc}

\usepackage{tabularx}
\usepackage{ragged2e} 
\newcolumntype{L}{>{\RaggedRight}X} 

\usepackage{comment}
\usepackage{makecell} 
\usepackage{multirow}

\begin{document}

\title{Lyapunov Optimization based Queue-aware Traffic Shaping for 5G-TSN in Industrial Environments}

\author{Kouros Zanbouri, Md Noor-A-Rahim, Cormac J Sreenan, Dirk Pesch\\
School of Computer Science and Information Technology\\ University College Cork, Ireland
}

\maketitle

\begin{abstract}
Manufacturing companies look increasingly at Private 5G networks to manage Automated Guided Vehicles (AGVs). While 5G promises Ultra-Reliable Low Latency Communication (URLLC), its service quality is challenged by industrial environments characterized by dense metallic structures, which frequently cause line-of-sight (LOS) blockage events, causing deep fades in received signal levels that can degrade channel capacity to near-zero. Standard transport protocols and rate adaptation mechanisms fail to react sufficiently fast to these deep fades, resulting in bufferbloat and latency spikes that violate safety margins. In this paper, we propose a cross-layer rate control algorithm based on Lyapunov Drift-plus-Penalty theory. The proposed controller dynamically optimizes the trade-off between service utility and queue stability based on instantaneous buffer states, without requiring predictive channel models. We validate the approach using a trace-driven simulation framework that replicates the stochastic dynamics of 5G blockage using 3GPP-compliant capacity data. Numerical results demonstrate that while baseline scheduling schemes suffer from catastrophic queue accumulation, leading to excessive delays upon reconnection, the proposed Lyapunov controller effectively eliminates bufferbloat. By preventing congestion-induced backlog, the system ensures immediate low-latency operation as soon as the channel recovers, maintaining near-deterministic behavior.
\end{abstract}



\section{Introduction}
Industry 4.0 relies heavily on the deployment of dependable wireless networks, such as private 5G networks, to support mobile robots and Automated Guided Vehicles (AGVs). Unlike traditional factory automation, these mobile agents require a wireless network infrastructure that handles mixed-criticality traffic. Private 5G networks with their Enhanced Mobile Broadband (eMBB) to support high-throughput video streaming and Ultra-Reliable Low Latency Communication (URLLC) to guarantee the responsiveness required for teleoperation and safety monitoring of AGVs~\cite{s21072489, 9247159} is the outstanding candidate technology.

To provide safe remote control, these networks must meet URLLC criteria, necessitating minimal latencies. However, many industrial radio environments are challenging with the presence of large metallic structures, moving cranes, and high-density racking systems, creating a complex propagation environment characterized by severe multipath fading and shadowing \cite{10123922}. In particular, when operating in mmWave or mid-band 5G frequencies, networked AGVs are susceptible to frequent line-of-sight (LOS) blockage events, where the channel capacity can degrade to near-zero very quickly \cite{9417520}.
To seamlessly integrate these wireless capabilities with deterministic industrial control systems, 3GPP Release 16 introduced the 5G-TSN framework, which logically models the entire 5G system as a single Time-Sensitive Networking (TSN) bridge \cite{3gpp23501}. At the network boundaries, time-synchronization and protocol translation are handled by two standard components: the Network-Side TSN Translator (NW-TT) at the core network, and the Device-Side TSN Translator (DS-TT) co-located with the User Equipment (UE). In the standard 3GPP architecture, the DS-TT functions as a standard Layer 2 bridge that, while performing residence-time computation for gPTP synchronization, lacks any awareness of the wireless channel state or queue conditions within the 5G modem, and therefore cannot adapt its forwarding behavior to the underlying radio dynamics.

A critical challenge arises when physical blockage events interact with this passive architecture and higher-layer protocols. Standard video transmission mechanisms (e.g., UDP/RTP) and congestion control algorithms (e.g., CUBIC TCP) rely on packet loss or Round Trip Time (RTT) signals to adapt their transmission rates \cite{9221188}. In a blockage scenario, these feedback loops are often too slow. As the channel capacity drops, the application layer continues to push high-bitrate video frames into the modem’s transmission buffer. 
Commercial 5G modems typically feature deep buffers (e.g., 4 to 8 MB) designed to absorb bursty traffic and maximize TCP throughput. For an 8 Mbps video stream, saturating such a buffer introduces multiple seconds of queuing latency, which can be catastrophic for real-time teleoperation safety margins. 

The mismatch between constant data generation and stochastic channel capacity inevitably results in bufferbloat at the transmission buffer \cite{11078615}. In the event of sustained blockages, this queuing delay can grow from milliseconds to seconds, far exceeding the operational bounds of real-time protocols. In safety-critical teleoperation scenarios, such latency renders remote machinery uncontrollable, posing an unacceptable risk to infrastructure and human workers alike.

To address this, we propose a queue-aware rate control algorithm. Our specific contributions are:
\begin{enumerate}
    \item \textbf{A Novel Architectural Placement (Smart DS-TT):} Design of a cross-layer control framework that embeds rate adaptation directly at the 5G-TSN device-side translator. This uniquely bridges application-layer video adaptation with instantaneous link-layer queue status, avoiding the high signaling overhead and slow reaction times of traditional end-to-end transport protocols.
    \item \textbf{Cross-Layer URLLC/eMBB Isolation:} Formulation of the AGV uplink rate control as a stochastic optimization problem using Lyapunov Drift-plus-Penalty theory. By dynamically throttling eMBB video traffic based on MAC-layer RLC queue occupancy, the system prevents video bufferbloat from suffocating co-scheduled URLLC-based AGV control traffic during deep fades.
    \item \textbf{Trace-Driven Validation of Deterministic Recovery:} Comprehensive statistical validation utilizing 3GPP-compliant capacity data. The results demonstrate that the proposed method mathematically bounds queuing delay, eliminates packet loss, and drastically reduces congestion-induced latency compared to standard baselines, ensuring immediate real-time operation upon channel recovery.
\end{enumerate}

\color{black}

\section{Related Work}
\label{sec:related_work}

Standard 3GPP mechanisms like Logical Channel Prioritization (LCP) \cite{3gpp38321} enforce QoS at the link layer but cannot throttle application source rates during sudden channel degradations. Conversely, end-to-end Adaptive Bitrate (ABR) frameworks including WebRTC and existing Lyapunov-based video streaming algorithms \cite{8550755} rely on Receiver Reports (e.g., RTCP) to estimate network conditions. This introduces an unavoidable RTT delay. During sudden 5G RF blockages, this delayed feedback can cause massive bufferbloat before the congestion signal ever reaches the sender \cite{9221188}. \color{black}

To manage bufferbloat closer to the radio edge, Active Queue Management (AQM) mechanisms like CoDel have been evaluated within the 5G SDAP layer. However, CoDel relies on static timing heuristics that fail to adapt to abrupt 5G bandwidth shifts, introducing throughput variance and risking unnecessary packet loss across flows in a shared queue. These methods also assume ideal traffic segregation, scaling poorly when diverse services share a single QoS Flow Indicator (QFI). Furthermore, CoDel’s congestion control mechanism depends on TCP’s window reduction response to packet loss. For UDP-based applications, dropped packets do not trigger any rate adaptation at the source, rendering the queue management ineffective as a throughput control tool \cite{8802027}.

Recent cross-layer architectures attempt to classify physical-layer blockages to adjust transport-layer behavior. Some rely on network-assisted proxies or base-station coordination~\cite{10154364}, deploying in-network caching and cross-layer pull mechanisms at the base station to work around TCP's slow recovery, but at the cost of infrastructure modifications and deployment overhead. Others operate end-to-end by tracking Signal-to-Interference-Plus-Noise Ratio (SINR) to suppress spurious Retransmission Timeouts (RTOs) or cap the congestion window~\cite{11132323}, but require custom protocol modifications at both the mobile station and the remote server, limiting practical deployability. Critically, these transport-layer mechanisms are designed exclusively for TCP flows, but fail for UDP-based real-time applications such as video streaming or sensor telemetry, where congestion window manipulation and RTO suppression are not available~\cite{11132323}.

Our work differentiates itself by utilizing a lightweight Lyapunov control law that requires only local buffer state information. By dynamically adapting the source rate based strictly on queue stability, our method avoids the rigid parameter tuning of AQM and the signaling overhead of complex network-assisted scheduling. This provides mathematically grounded delay bounds without sacrificing throughput, regardless of the underlying transport protocol.

\section{Proposed 5G-TSN Smart DS-TT Architecture}
\label{sec:system_architecture}
\begin{figure}[t]
    \centering
    \begin{tikzpicture}[
        node distance=0.6cm,
        box/.style={draw, rectangle, text width=6.5cm, minimum height=0.8cm, align=center, font=\small\sffamily, thick},
        line/.style={thick}
    ]
        
        \node[box, fill=gray!10] (cnc) {\textbf{CNC} \\ (Centralized Network Configuration)};
        
        \node[box, fill=blue!10, below=of cnc, yshift=-0.2cm] (tsnaf) {\textbf{TSN AF} \\ (TSN Application Function)};
        
        \node[box, fill=blue!10, below=of tsnaf] (nwtt) {\textbf{NW-TT} co-located with \textbf{UPF} \\ (Network-Side TSN Translator)};
        
        \node[box, fill=blue!10, below=of nwtt, yshift=-0.2cm] (gnb) {\textbf{5G RAN} \\ (gNodeB)};
        
        \node[box, fill=blue!10, below=of gnb, yshift=-0.2cm] (dstt) {\textbf{DS-TT} co-located with \textbf{UE} \\ (Device-Side TSN Translator)};
        
        \node[box, fill=gray!10, below=of dstt, yshift=-0.2cm] (agv) {\textbf{AGV Application Layer} \\ (End-Device / Sensors \& Actuators)};

        \draw[line] (cnc) -- (tsnaf);
        \draw[line] (tsnaf) -- node[right, font=\footnotesize] {Control Plane} (nwtt);
        \draw[line] (nwtt) -- node[right, font=\footnotesize] {User Plane (N3)} (gnb);
        \draw[line] (gnb) -- node[right, font=\footnotesize] {Uu Air Interface} (dstt);
        \draw[line] (dstt) -- node[right, font=\footnotesize] {Standard Ethernet} (agv);

        \draw[dashed, thick, blue] ($(tsnaf.north west)+(-0.2,0.3)$) rectangle ($(dstt.south east)+(0.6,-0.3)$);
        \node[blue, right, font=\footnotesize\bfseries, align=center] at ($(nwtt.east)!0.5!(gnb.east)+(0.6,0)$) {\rotatebox{-90}{5G System (Logical Bridge)}};

        \draw[dotted, thick, green!50!black] ($(dstt.north west)+(-0.6,0.25)$) rectangle ($(agv.south east)+(0.2,-0.3)$);
        \node[green!50!black, left, font=\footnotesize\bfseries, align=center] at ($(dstt.west)!0.5!(agv.west)+(-0.6,0)$) {\rotatebox{90}{Physical AGV}};

    \end{tikzpicture}
    \caption{Overview of the 3GPP Rel-16 5G-TSN architecture.}
    \label{fig:5gtsn_overview}
\end{figure}
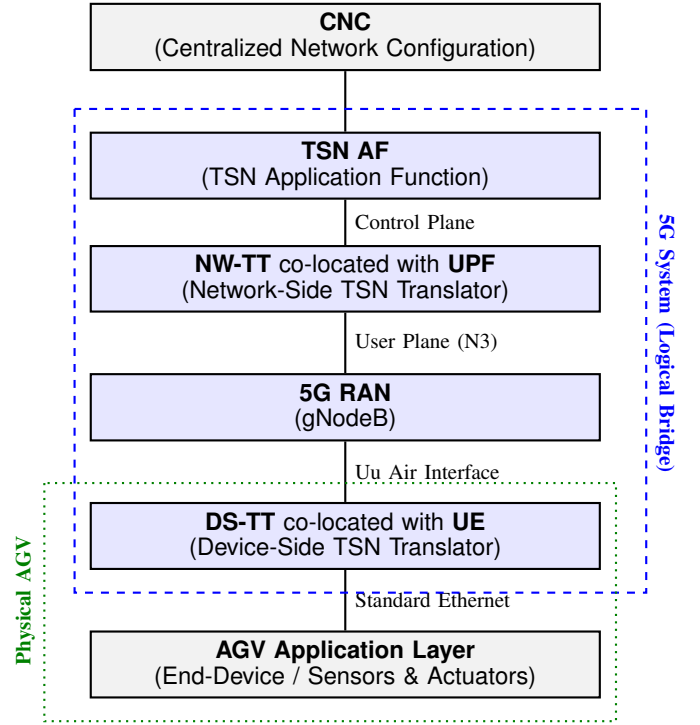

As illustrated in Fig.~\ref{fig:5gtsn_overview}, the 3GPP Release 16 framework bridges the gap between deterministic wired domains by modeling the entire cellular infrastructure as a transparent node. The 5G System logically acts as a single IEEE 802.1 bridge (blue dashed box). The DS-TT serves as the critical intersection point, logically integrating with the 5G network while physically co-located on the industrial AGV (green dotted box). However, because this standard DS-TT operates purely as a passive forwarding interface, it remains completely blind to the instantaneous state of the wireless channel.

\begin{figure*}[ht]
    \centering
    \includegraphics[width=0.7\textwidth]{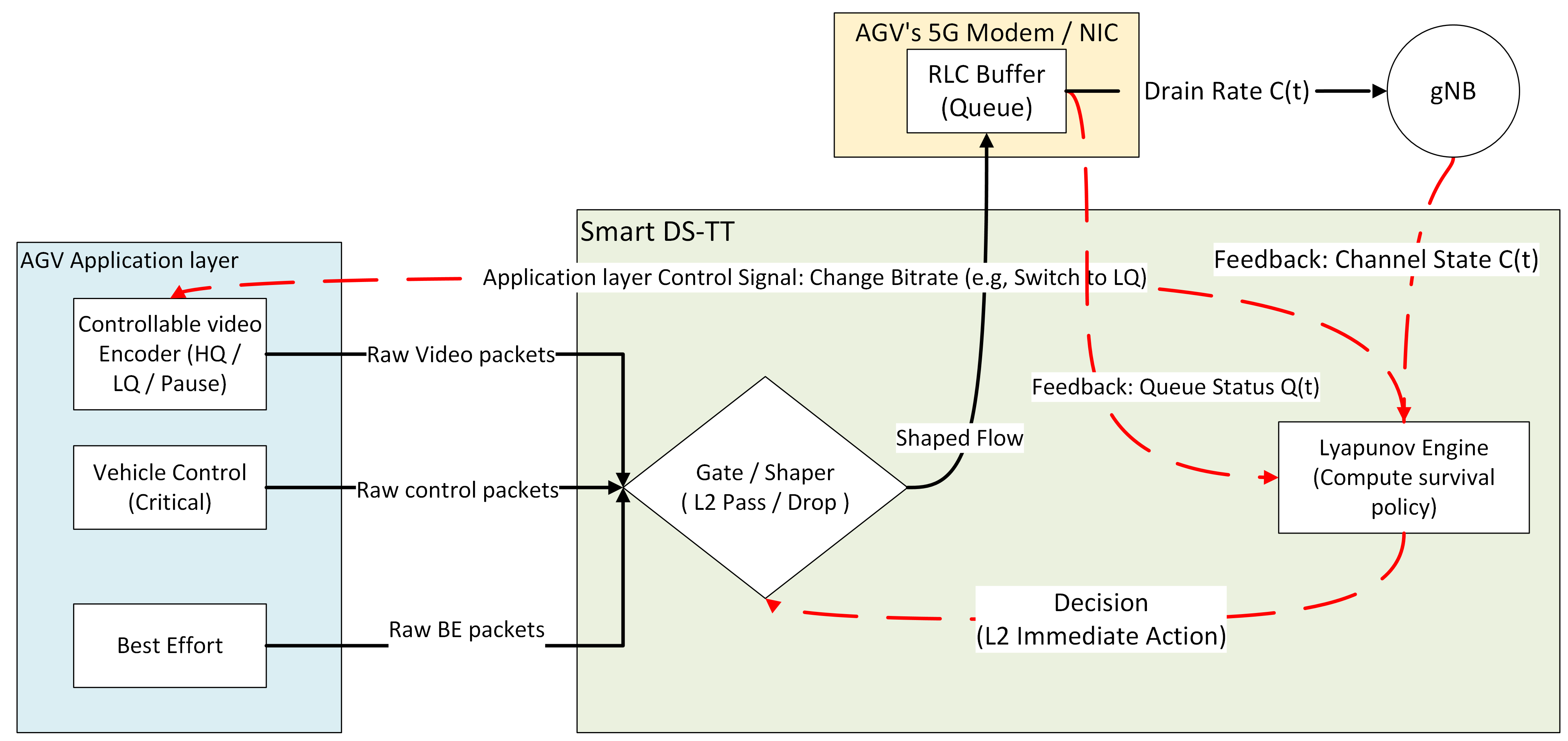}
    \caption{System Overview: The proposed Smart DS-TT architecture}
    \label{fig:smart-dstt:system_overview}
\end{figure*}

To resolve this architectural vulnerability, we propose a novel ``Smart DS-TT'' architecture. Rather than acting as a blind Layer 2 bridge between the AGV and the 5G modem, our design elevates the DS-TT into an active, cross-layer control node governed by Lyapunov optimization. The proposed system is modeled as three distinct interacting domains: the Application Layer, the Smart DS-TT, and the 5G Radio Access Network (RAN), as detailed in Fig.~\ref{fig:smart-dstt:system_overview}.

\subsection{Application Traffic Model \& 5G QoS Mapping}
The proposed Smart DS-TT operates at the logical boundary between the AGV's internal network and the 5G System (5GS). In the context of 3GPP Release 16 TSN integration, our controller functions as an enhanced \textbf{Device-Side TSN Translator (DS-TT)}. Instead of relying solely on the network-side NW-TT for policing, our Smart DS-TT implements strict uplink traffic shaping at the UE to align with standard 5G Quality of Service (QoS) flows. 
The AGV generates three distinct classes of traffic, which the Smart DS-TT maps and manages as follows:

\begin{enumerate}
    \item[\textbf{Critical Control Traffic (URLLC)}] flows consist of periodic safety messages (e.g., heartbeat signals, emergency stop commands). 
    These packets are mapped to a high-priority 5QI (e.g., 5QI 82) with a strict Packet Delay Budget (PDB) of $<10$ ms. The Smart DS-TT ensures this queue is always emptied first using a Strict Priority scheduling discipline.
    
    \item[\textbf{Adaptive Video Streaming (eMBB):}] This flow provides environmental awareness for teleoperation. Unlike standard fixed-bitrate video, this source is controllable (switching between 1080p, 720p, 480p, or Pause). Packets are mapped to a Guaranteed Bit Rate (GBR) flow (e.g., 5QI 2). Crucially, the Lyapunov controller modulates the offered load to this flow. By ensuring the video generation rate never exceeds the instantaneous capacity supported by the channel, it actively prevents the 5G Modem's RLC buffer from overflowing.
    
    \item[\textbf{Best Effort Data (mMTC):}] This represents background traffic, such as log uploads or non-critical sensor telemetry. 
    This traffic is delay-tolerant, mapped to a default non-GBR flow (e.g., 5QI 9), and serves as the lowest priority queue.
\end{enumerate}

While modern AGVs operate autonomously, industrial safety protocols typically require a continuous "Passive Monitoring" video stream (low bitrate) to be maintained for liability logging and situational awareness. In the event of an exception (e.g., obstacle detection), a remote operator must instantly take control. If the video session were established only on-demand, the protocol handshake delay (RTSP/SIP setup) would violate safety margins. Therefore, we model a continuous video uplink that acts as a heartbeat, capable of ramping up to high resolution immediately upon teleoperation takeover.

Crucially, the 5G New Radio (NR) scheduler does not statically reserve the maximum video bandwidth during this passive autonomous phase; instead, unused radio resources are dynamically reallocated by the gNB to other users to maximize network efficiency. When an operator takes over, the sudden surge in high-definition uplink demand must contend within shared radio resources and operate with potentially hostile Industrial IoT channel shadowing. If the application blindly injects high-definition video before the physical channel capacity can fully accommodate the surge, catastrophic bufferbloat ensues. This dynamic bandwidth availability necessitates our cross-layer pre-shaping, which ensures that the 5G RAN scheduler is never overwhelmed by unyielding application traffic. By policing queues before they reach the modem, the system allows standard 5G QoS mechanisms to function correctly, even during deep channel blockages.
\color{black}
\subsection{The Smart DS-TT Architecture}
The core contribution of this work is the Smart DS-TT, which sits between the application layer and the 5G modem. It implements a cross-layer feedback mechanism to ensure survival during channel outages. The architecture consists of two primary functional blocks:

\subsubsection{The Lyapunov Engine}
This module executes the core rate-adaptation policy, as depicted in Fig.~\ref{fig:smart-dstt:system_overview}. Instead of relying on static QoS rules or delayed end-to-end transport metrics, it continuously observes the current Radio Link Control (RLC) video queue backlog $Q(t)$. We operate under the practical assumption that this queue state is exposed by the commercial 5G modem via standard or vendor-specific APIs, requiring no underlying protocol modifications (as detailed in Section~\ref{sec:feasibility}). While the instantaneous channel capacity $C(t)$ dictates the physical packet departure rate at the MAC layer, the Lyapunov engine relies strictly on $Q(t)$, which accumulates the physical mismatch between application arrivals and $C(t)$ as a highly accurate, localized proxy for network congestion. Using this queue state, the engine optimizes a Lyapunov drift-plus-penalty objective function to maximize video utility while guaranteeing queue stability. Working in tandem with the physical MAC layer, the overall Smart DS-TT framework enforces two distinct control actions:
\begin{enumerate}
    \item[\textbf{Application Layer Control:}] A discrete command computed by the Lyapunov engine and sent back to the Video Encoder to proactively shape the source generation rate (e.g., downgrade to 480p or Pause) before congestion propagates.
    \item[\textbf{Layer 2 Decision:}] An immediate gating action executed by the Traffic Shaper to prioritize critical URLLC packets and buffer non-essential eMBB traffic according to the available $C(t)$.
\end{enumerate}
\color{black}

\subsubsection{Gate / Shaper}
This component physically intercepts raw packets from the application layer. It implements a strict priority scheduling discipline with a \textbf{Freshness-First} policy for critical uplink traffic.
Unlike standard First-In-First-Out (FIFO) queuing, the URLLC buffer is optimized for the deadline-driven nature of industrial control applications. In such systems, control payloads represent highly time-sensitive machine state updates. If a packet's delivery deadline is exceeded, the information becomes obsolete; transmitting stale data not only provides no value to the application but actively wastes critical bandwidth. To reflect this, our architecture maintains the URLLC queue as a single-packet buffer for this specific telemetry application. New state updates immediately overwrite older ones, ensuring that obsolete packets are proactively discarded. This guarantees that the modem always transmits the freshest possible state upon channel recovery, further warranting the explicit architectural separation of URLLC control payloads from the continuous video flow. This ensures both
\begin{enumerate}
    \item[\textbf{Freshness:}] stale telemetry (e.g., position estimates or velocity readings past their delivery deadline) is discarded at source, minimizing Age of Information (AoI) and preventing control loop instability caused by the remote controller acting on obsolete state data.
    \item[\textbf{Priority:}] the Gate maintains strictly isolated queues and forwards payloads from the high-priority URLLC queue into the transmission buffer before servicing any Video or mMTC traffic. This strict isolation ensures that during a blockage event (where $C(t) \to 0$), critical safety signals (e.g., heartbeats or emergency stop messages) are transmitted immediately upon recovery, never blocked by a backlog of video data.
\end{enumerate}

\subsection{5G System and Feedback Loop}
The final component is the 5G Network Interface (Modem/NIC) and the gNB (gNodeB/Next-Generation Node B). Building upon the previously established architectural assumption of accessible modem buffer states, we treat the instantaneous RLC queue occupancy as the primary state variable for our control algorithm. The system dynamics are modeled using the standard discrete-time queue evolution equation \cite{Neely2010Stochastic}:
\begin{equation}
    Q(t+1) = \max[0, Q(t) - C(t)] + A(t)
    \label{eq:queue_dynamics}
\end{equation}
Where $Q(t)$ is the backlog of the modem's RLC transmission buffer (in bits), while $A(t)$ and $C(t)$ represent the amount of data arriving from the Smart DS-TT and the amount of data departing over the wireless channel during slot $t$, respectively. \color{black} The Smart DS-TT ensures that $A(t)$ is dynamically throttled whenever $C(t)$ drops, preventing $Q(t)$ from growing beyond the safety threshold required for URLLC traffic.
Consequently, we formulate the rate adaptation problem as a trade-off between maximizing video quality (Utility) and minimizing network congestion (Queue Stability) in the following.

\subsection{Lyapunov Engine} 
\label{sec:LyapunovEngine}

To formalize the trade-off between maximizing video quality and minimizing network delay, we first define our system's objective. At any time slot $t$, the system achieves a certain Quality of Experience (QoE), which we quantify using a utility function $U(t)$. Because human perception of video quality exhibits diminishing returns at higher bit rates, we model $U(t)$ using a profile-specific discrete utility weight $w_p$ pre-assigned to each profile (as defined in Table~\ref{tab:profiles}):
\begin{equation}
    U(t) = w_p, \quad p = \text{selected profile at time } t
\end{equation}

By mapping the discrete video profiles to a monotonically increasing utility set ($w_{1080p} > w_{720p} > w_{480p} > w_{Pause}$), this formulation approximates the diminishing returns of human visual perception while remaining compatible with the drift-plus-penalty framework. \color{black}
Our goal is to maximize the time-averaged utility while keeping the network queue stable. Recall from the queue evolution model in Eq.~\eqref{eq:queue_dynamics} that the backlog $Q(t)$ is driven by the arrival rate $A(t)$, which is directly determined by our chosen video transmission rate $R_{video}(t)$, and depleted by the uncontrollable channel capacity $C(t)$. 
To mathematically measure and penalize this accumulated congestion, we define a quadratic Lyapunov function \cite{9895362}:
\begin{equation}
    L(t) = \frac{1}{2} Q(t)^2
\end{equation}
The one-step Lyapunov drift, defined as $\Delta(t) = L(t+1) - L(t)$, represents the expected change in congestion over one time slot. Squaring the queue evolution equation yields the fundamental bound on this drift:
\begin{equation}
    \Delta(t) \le B + Q(t)(A(t) - C(t))
\end{equation}
where $B$ is a positive bounding constant satisfying $B \ge \frac{1}{2}\bigl(A(t)^2 + C(t)^2\bigr)$ for all $t$, which follows from applying the inequality $\max(x,0)^{2} \le x^{2}$ to the queue evolution and bounding the resulting squared arrival and departure terms.

Following the drift-plus-penalty framework \cite{9152999, ZHAO2024103580}, we incorporate our utility function to form the drift-plus-penalty expression, $\Delta(t) - V \cdot U(t)$. By substituting our drift bound, we obtain the upper bound for our optimization objective:
\begin{equation}
    \Delta(t) - V \cdot w_p \le B - V \cdot w_p + Q(t)\bigl(R_p - C(t)\bigr)
\end{equation}
where $V$ is a non-negative control parameter that dictates the strictness of the trade-off: a larger $V$ greedily prioritizes Utility (higher video rates), while a smaller $V$ prioritizes queue stability (lower latency). 

Because the video arrival rate $A(t)$ is directly controlled by the selected profile $R_{video}(t)$, and $C(t)$ is independent of our control actions, minimizing the right-hand side of this bound is mathematically equivalent to maximizing the following metric at each time slot $t$:
\begin{equation}
    \underset{p \,\in\, \mathcal{P}}{\text{maximize}}
    \quad V \cdot w_p \;-\; Q(t) \cdot R_p
    \label{eq:optimization_metric}
\end{equation}

where $w_p$ is the utility weight and $R_p$ is the transmission rate (Mbps) of profile $p$, both drawn from Table~\ref{tab:profiles}. To ensure dimensional consistency and prevent the penalty term from mathematically dominating the utility, $Q(t)$ is explicitly normalized to Megabits (Mb) and $R_p$ is evaluated in Megabits per second (Mbps). This metric is evaluated for each profile at every TTI, and the profile maximising it is selected as the control action. In this final policy, $Q(t)$ acts as a dynamic penalty weight. By continuously observing the current queue length $Q(t)$, the engine selects the video rate $R_{video}(t)$ that maximizes this metric, effectively preventing bufferbloat by throttling the data rate before congestion can accumulate. While the instantaneous channel capacity $C(t)$ is not directly present in this decision metric, the queue backlog $Q(t)$ acts as the historical integral of the channel state, ensuring the controller implicitly reacts to capacity degradations.

Furthermore, this formulation yields deterministic, mathematically derivable profile-switching thresholds. By setting the decision metrics of two profiles equal, we can determine the exact queue backlog that triggers a downgrade. For example, using the values from Table~\ref{tab:profiles}, the system will downgrade from 1080p to 720p when $V(8.0) - Q(t)(8.0) = V(5.0) - Q(t)(4.0)$, which simplifies to $Q(t) = 0.75V$. This proves that the system reacts proactively to queue accumulation before absolute physical saturation occurs.
\color{black}

The performance of the drift-plus-penalty algorithm is strictly governed by the control parameter $V$. This parameter adjusts how the system prioritizes queue stability through backlog minimization versus achieving higher video bitrates.
\begin{enumerate}
    \item[\textbf{Small $V$ (Safety Focus):}] The $Q(t) \cdot R$ term dominates. The controller becomes highly sensitive to even small queue build-ups, aggressively throttling the video generation rate (or pausing it entirely) to keep $Q(t) \approx 0$. This ensures strict adherence to minimal latency but may result in frequent quality downshifts or unnecessary video suppression during minor channel dips.
    \item[\textbf{Large $V$ (Quality Focus):}] The $V \cdot U$ term dominates. The controller tolerates a larger queue backlog $Q(t)$ before the cost of queuing outweighs the reward of high-quality video. This yields higher average video quality but increases the risk of latency spikes.
\end{enumerate}

For industrial AGV teleoperation, latency is a hard constraint. Therefore, $V$ must be selected such that the maximum steady-state queue size $Q_{max}$ does not induce a delay exceeding the safety threshold $\tau_{max}$ \cite{itu_g1070_2018}. 
Algorithm \ref{alg:lyapunov_control} summarizes the operational logic of the Smart DS-TT. Several practical advantages emerge from this cross-layer design. First, the computational complexity is extremely low, scaling linearly with the number of video profiles ($O(|\mathcal{P}|)$) rather than the number of network flows. On the UE side, the DS-TT only manages a small, fixed set of profiles for a single AGV (e.g., four discrete rates); thus, the decision logic introduces negligible processing delay, allowing it to execute strictly within the 0.5 ms TTI limit. Furthermore, this linear complexity ensures the approach is highly scalable and could readily be adapted for downlink traffic management within NW-TT, which might need to simultaneously handle thousands of profiles for an entire fleet. Finally, because the control law is deterministic and relies solely on instantaneous feedback, the system remains robust to stochastic channel fluctuations without requiring computationally expensive channel prediction or training-heavy machine learning models.

As detailed in Algorithm 1, the optimal video bitrate is computed prior to accounting for the instantaneous URLLC reservation. While a strictly coupled formulation might attempt to subtract the expected URLLC arrival rate, industrial URLLC payloads (e.g., 50–-100 byte heartbeats) are dimensionally negligible compared to Megabit-scale video flows. Furthermore, because the Smart DS-TT operates in a distributed manner exclusively at the UE edge, cell-wide URLLC congestion from multiple AGVs is inherently reflected in the reduced uplink capacity grant $C(t)$ assigned by the gNB, meaning the local algorithm scales without requiring cross-UE orchestration. Therefore, ignoring the local URLLC footprint in the application layer Lyapunov decision preserves the algorithm's computational simplicity without sacrificing optimality.
\color{black}

\section{Trace-Driven Simulation}
\label{sec:Simulation}
\subsection{Experimental Setup}
To validate the proposed Smart DS-TT algorithm without the safety risks associated with physical testing on active industrial machinery, we developed a discrete-time, trace-driven simulation framework. This framework models the stochastic interaction between wireless channel dynamics and the AGV's internal queuing system, specifically focusing on the catastrophic impact of channel blockages.
The simulation compares the proposed Smart DS-TT agent against a Baseline UDP Agent. The Baseline represents a standard industrial camera feed that attempts to transmit HQ video regardless of channel conditions, mimicking current non-adaptive deployments. The Proposed Lyapunov engine utilizes the Lyapunov drift-plus-penalty algorithm defined in Section \ref{sec:LyapunovEngine} to dynamically adjust the video encoding rate $R(t)$ based on the instantaneous queue backlog. The specific parameters utilized in the experiment are detailed in Table \ref{tab:sim_params}. The Utility weights used in this work are listed in Table~\ref{tab:profiles}. For computational efficiency, $w_p$ is pre-computed offline for each profile and stored as a constant, since the profile set $\mathcal{P}$ is fixed.

\begin{algorithm}
\caption{Smart DS-TT Cross-Layer Control Logic}
\label{alg:lyapunov_control}
\begin{algorithmic}[1]
\Require Tuning parameter $V$, Time slot $\Delta t$
\Require Video Profiles $\mathcal{P} = \{(R_p, w_p)\}$ as per Table~\ref{tab:profiles}
\State \textbf{Initialize:} $Q_{video} \gets 0$, $Q_{urllc} \gets 0$, $Q_{mmtc} \gets 0$

\Loop { for each TTI $t = 1, 2, \dots$}

    \State \textbf{// Step 1: Lyapunov Control Decision (Application Layer)}
    \State $R^* \gets 0$
    \State $D_{max} \gets -\infty$
    
    \ForAll {profile $p \in \mathcal{P}$}
        \State $R_p \gets p.rate$
        \State $w_p \gets p.Utility\_weight$ 
        \Comment{Utility Weight from Table~\ref{tab:profiles}}
        \State \textbf{Compute Drift-plus-Penalty metric:}
        \State $\delta \gets V \cdot w_p - Q_{video}(t) \cdot R_p$
        \Comment{Eq.~\eqref{eq:optimization_metric}}
        
        \If {$\delta > D_{max}$}
            \State $D_{max} \gets \delta$
            \State $R^* \gets R_p$ \Comment{Select optimal bitrate}
        \EndIf
    \EndFor
    \State \textbf{Action:} Send $R^*$ command to Video Encoder

    \State \textbf{// Step 2: Traffic Arrival (Input)}
    \State $Q_{video} \gets Q_{video} + R^* \cdot \Delta t$
    \If {URLLC period arrived}
        \State $Q_{urllc} \gets \text{Payload}_{URLLC}$ \Comment{Freshness update}
    \EndIf
    \State $Q_{mmtc} \gets Q_{mmtc} + R_{mmtc} \cdot \Delta t$

    \State \textbf{// Step 3: Priority Scheduling (Layer 2)}
    \State Get Channel Capacity $C(t)$ from 5G Modem
    \State $C_{rem} \gets C(t)$
    
    \State \Comment{Priority 1: Critical Control}
    \State $D_{u} \gets \min(Q_{urllc}, C_{rem})$
    \State $Q_{urllc} \gets Q_{urllc} - D_{u}$
    \State $C_{rem} \gets C_{rem} - D_{u}$
    
    \State \Comment{Priority 2: Adaptive Video}
    \State $D_{v} \gets \min(Q_{video}, C_{rem})$
    \State $Q_{video} \gets Q_{video} - D_{v}$
    \State $C_{rem} \gets C_{rem} - D_{v}$
    
    \State \Comment{Priority 3: Best Effort}
    \State $D_{m} \gets \min(Q_{mmtc}, C_{rem})$
    \State $Q_{mmtc} \gets Q_{mmtc} - D_{m}$

\EndLoop
\end{algorithmic}
\end{algorithm}

\subsection{Parameter Tuning \& Sensitivity}
\label{sec:Simulation_P_Tunning}
A critical aspect of the proposed framework is the tuning of the Lyapunov control parameter $V$, which governs the trade-off between maximizing video utility (QoE) and minimizing queue backlog. Theoretically, the maximum queue backlog is bounded by $O(V)$. 
To select the optimal $V$, we perform an offline calibration phase (shown in Fig. \ref{fig:v_tradeoff}). We perform a logarithmic sweep of $V$ across 25 points over $[0.1, 100]$ to identify the system's operating regions. Based on the empirical results detailed in Section \ref{sec:Results}-A, we selected \textbf{$V = 0.5$}, which provides a sweet spot that maintains High Quality (1080p) video during healthy channel conditions while triggering an immediate reaction when the queue backlog exceeds a safety threshold of $\approx 1$ Mbit.

\begin{table}[htbp]
\caption{Simulation Parameters}
\begin{center}
\begin{tabular}{|l|l|}
\hline
\textbf{Parameter} & \textbf{Value} \\
\hline
Channel Model & 3GPP Numerology 1 (FR1) \\
\hline
TTI Duration & 0.5 ms \\
\hline
Blockage Scenarios & 2s (transient), 5s (deep), 3s partial (2\,Mbps) \\
\hline
Avg. Link Capacity & $\approx$ 15 Mbps (Unblocked) \\
\hline
Buffer Size Limit & Finite (evaluated) vs. Infinite \\
\hline
Finite Buffer Ceiling & 4 Mbits \\
\hline
Lyapunov Parameter ($V$) & Swept ($0.1$ to $100$) \\
\hline
Traffic Classes & URLLC (Critical) + Video (Adaptive) + BE \\
\hline
mMTC Background Rate & 2 Mbps \\
\hline
\end{tabular}
\label{tab:sim_params}
\end{center}
\end{table}

\vspace{-0.5cm}

\begin{table}[htbp]
\caption{Video Profile Parameters}
\begin{center}
\begin{tabular}{|l|c|c|}
\hline
\textbf{Profile} & \textbf{Rate (Mbps)} & \textbf{Utility Weight $w_p$} \\
\hline
1080p & 8.0 & 8.0 \\
\hline
720p  & 4.0 & 5.0 \\
\hline
480p  & 1.0 & 2.0 \\
\hline
Pause & 0.0 & 0.0 \\
\hline
\end{tabular}
\label{tab:profiles}
\end{center}
\end{table}

\vspace{-0.25cm}

\subsection{Channel Emulation \& Blockage Scenarios}
\label{sec:ChannelSimulation}
To bridge the gap between theoretical modeling and realistic channel dynamics, we utilized the \textbf{Simu5G} framework within OMNeT++ to generate high-fidelity synthetic capacity traces. While our control architecture is fundamentally trace-agnostic and could readily ingest empirical capacity data from a live 5G deployment, such high-resolution, blockage-specific industrial datasets are not currently publicly available. Consequently, to evaluate the system under controlled and reproducible conditions, the simulation parameters were configured to generate a synthetic representation of a harsh industrial environment:
\begin{enumerate}
    \item[\textbf{Scenario:}] 3GPP Indoor Factory (InF-DH) channel model with dense clutter (high fading variance).
    \item[\textbf{Link Direction:}] Uplink (UL) only, modeling the video stream from the AGV to the gNB.
    \item[\textbf{Mobility:}] The AGV moves at 2 m/s along linear trajectories representing factory aisles, encountering periodic shadowing from metallic racks.
    \item[\textbf{MAC/PHY Layer:}] The gNB utilizes a \textit{Proportional Fair} scheduler with $\mu=1$ (Numerology 1). The 'Capacity' values in the trace represent the instantaneous Transport Block Size (TBS) allocated to the UE per TTI, accounting for HARQ retransmissions and MCS adaptation.
\end{enumerate}
These traces capture the rapid fluctuations of the SINR and the resulting transport block availability, serving as the ground truth input $C(t)$ for our discrete-time evaluation.

To rigorously evaluate the system's resilience, we design scenarios that reflect the harsh propagation realities of metal-dense industrial environments. We introduce the following Blockage Scenarios:

\begin{enumerate}

    \item[\textbf{Transient Deep Blockage (2s):}] Represents dynamic metallic obstacles, such as passing forklifts. 
    
    \item[\textbf{Extended Deep Blockage (5s):}] Represents the AGV navigating behind large static machinery or turning into a shadowed, metal-racked aisle.

     \item[\textbf{Severe Capacity Degradation (Partial Blockage-3s):}] Represents a passing human or non-metallic obstacle. 
    
\end{enumerate}

For scenarios 1 and 2, the simulated channel capacity drops to $0$ Mbps. While a complete $0$ Mbps outage may seem extreme in standard macro-cellular Sub-6 GHz (FR1) networks, it is a very realistic phenomenon in dense Industrial Indoor Factory (InF) deployments. The presence of moving metallic cranes and high-density racking systems causes severe Non-Line-of-Sight (NLOS) shadowing. During these deep impairment events, the SINR frequently drops below the minimum threshold required to decode the lowest Modulation and Coding Scheme (MCS), resulting in transient zero-capacity outages. Furthermore, from a control-theoretic perspective, evaluating the system under a step-function drop to $0$ Mbps serves as a strict worst-case boundary test. Demonstrating duration-independent Lyapunov queue stability during absolute zero-capacity events guarantees that the controller will mathematically survive any lesser, partial degradation.
\color{black}

\section{Simulation Results \& Discussion}
\label{sec:Results}

\subsection{Impact of Lyapunov Parameter $V$} 
\begin{figure}[htbp]
\centering
\includegraphics[width=\columnwidth]{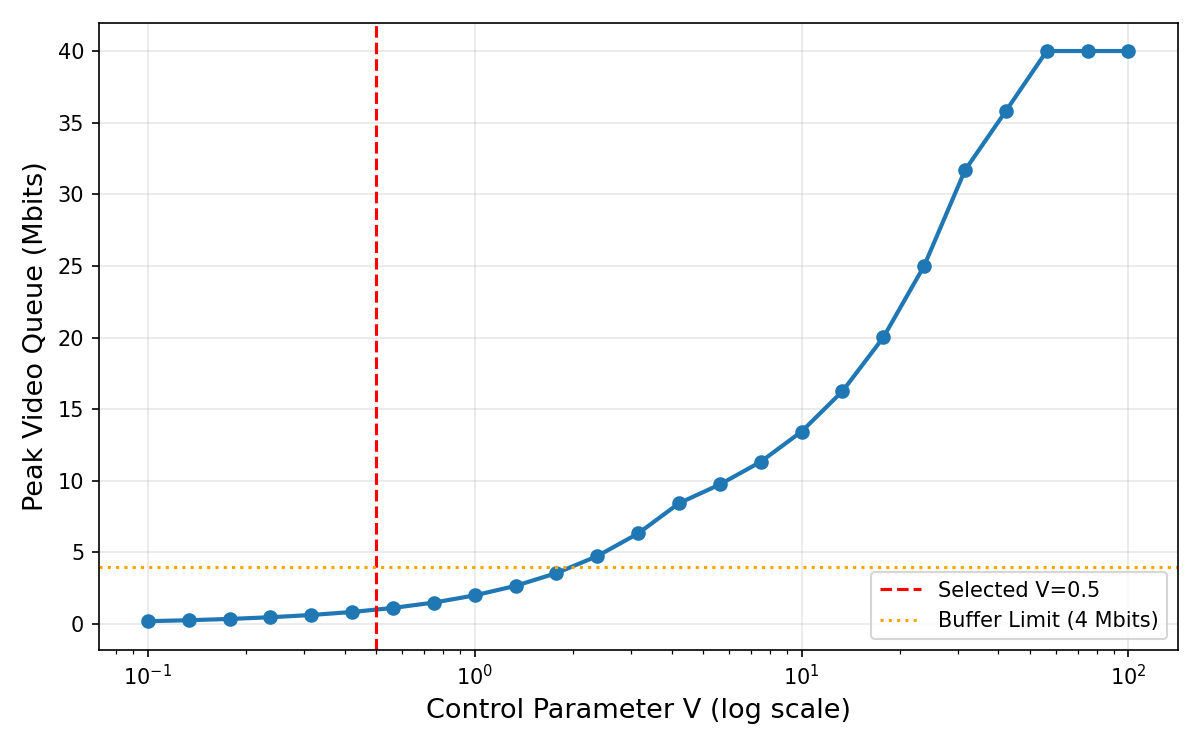}
\caption{Impact of V on Queue Stability (5s Blockage)}
\label{fig:v_tradeoff}
\end{figure}

Fig.~\ref{fig:v_tradeoff} illustrates the peak video queue accumulated during a 5-second blockage event as $V$ is swept logarithmically across 25 points from $0.1$ to $100$. The results reveal three distinct operating regions:

\subsubsection{The Linear Control Region} 
For $V \le 2.37$, the peak queue follows a precise linear relationship:
\begin{equation}
    Q_{peak} \approx 2V \quad \text{(Mbits)}
    \label{eq:qpeak_linear}
\end{equation}
which confirms the theoretical $O(V)$ bound predicted by the drift-plus-penalty framework~\cite{Neely2010Stochastic}. Within this region, the controller maintains deterministic and predictable queue behaviour, a critical property for safety-certified industrial deployments.

At our selected operating point of \textbf{$V = 0.5$}, 
equation~\eqref{eq:qpeak_linear} predicts $Q_{peak} = 1.00$\,Mbit, confirmed exactly by simulation. This corresponds to a worst-case video queuing delay of:
\begin{equation}
    D_{max} = \frac{Q_{peak}}{\bar{C}} 
            = \frac{1.00\,\text{Mb}}{15\,\text{Mbps}} 
            \approx 66.7\,\text{ms}
\end{equation}
This easily satisfies standard teleoperation latency targets, while the co-scheduled URLLC queue independently maintains sub-1.5\,ms queuing delay during active channel states through strict priority scheduling.

\subsubsection{The Transition Region}
For ($2.37 < V < 56.2$), the 4\,Mbit physical buffer ceiling is first exceeded and the linear relationship breaks down. The peak queue grows non-linearly with $V$, reflecting the physical constraint imposed by the maximum video generation rate. Within this region, the controller still provides partial queue suppression relative to the Baseline, but no longer offers deterministic delay guarantees and cannot be safely deployed in latency-sensitive industrial scenarios.

\subsubsection{The Saturation Region}
For $V \ge 56.2$, the controller degenerates into Baseline behaviour. The queue saturates at:
\begin{equation}
    Q_{sat} = R_{max} \times T_{block} 
            = 8\,\text{Mbps} \times 5\,\text{s} 
            = 40\,\text{Mb}
\end{equation}
confirming that at high $V$ the utility term completely dominates and the agent transmits at maximum rate throughout the blockage, identical to the uncontrolled Baseline. This saturation value validates the physical consistency of the simulation.
Our selection of $V = 0.5$ is therefore analytically derived, empirically confirmed, and places the system deep within the linear control region where queue behaviour is fully predictable.

\subsection{Resilience to Variable Blockage Events}
\begin{figure}[htbp]
\centering
\includegraphics[width=0.97\columnwidth]{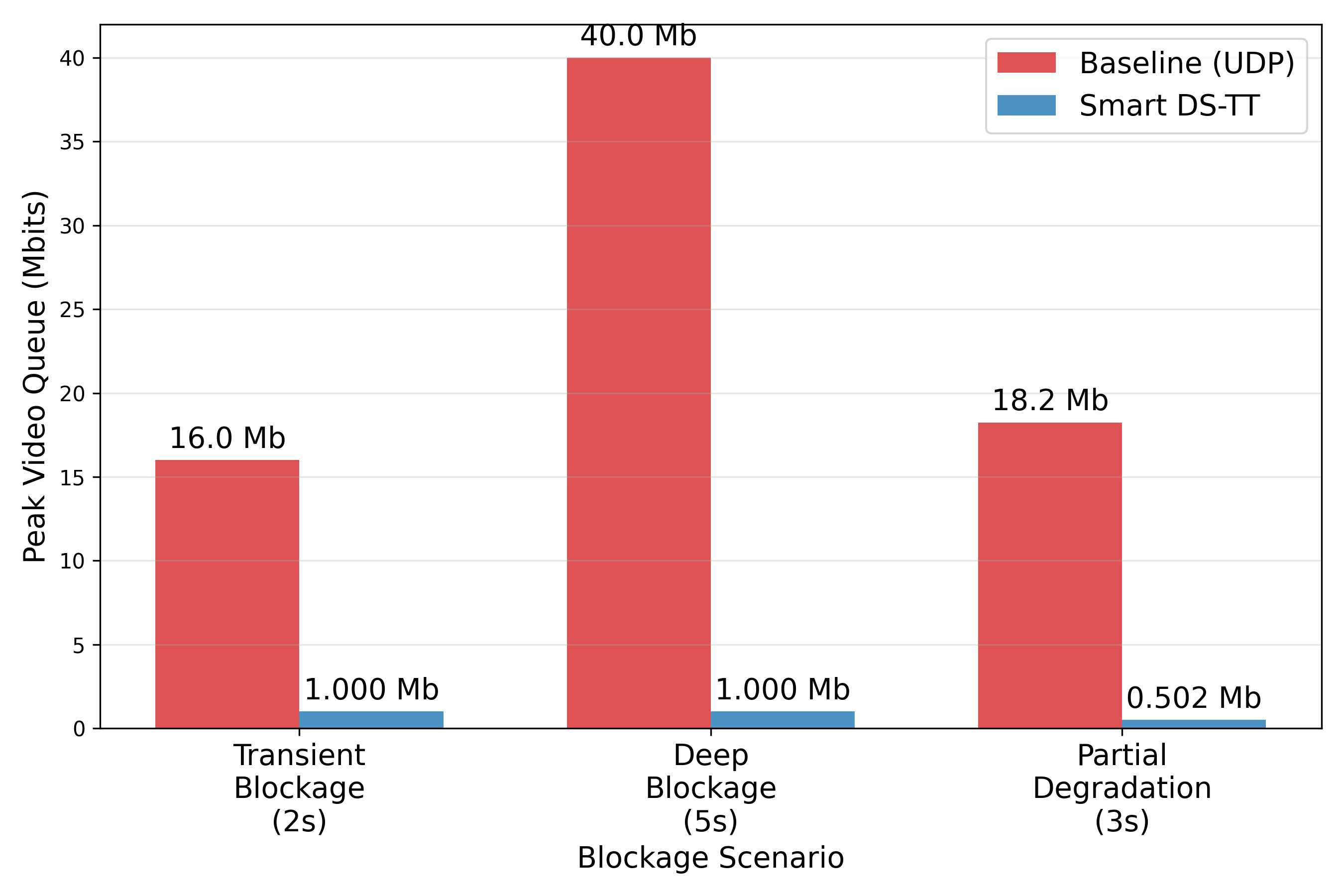}
\caption{Peak video queue comparison across three realistic industrial blockage scenarios}
\label{fig:realistic_scenario}
\end{figure}

We evaluate controller resilience using the realistic multi-scenario channel trace generated and described in Sec .~\ref {sec:ChannelSimulation}, which contains three distinct impairment windows representative of different industrial obstacle types. Both agents are evaluated on the identical trace, ensuring a fair comparison under the same channel conditions. Fig.~\ref{fig:realistic_scenario} shows the peak video queue accumulated by each agent across all three scenarios.

\subsubsection{Transient Blockage (starting at 5\,s, lasting 2\,s)}
This scenario models a transient blockage caused by a passing obstacle interrupting the line-of-sight path, resulting in a complete outage (0\,Mbps) for 2\,s. Given the AGV speed of 2\,m/s, this corresponds to a blockage over approximately 4\,m, which is representative of larger movable obstacles (e.g., forklifts).

The Baseline agent continues transmitting at 8\,Mbps throughout, accumulating a peak queue of 16.00\,Mb exactly matching the theoretical deficit of $R_{max} \times T_{block} = 8\,\text{Mbps} \times 2\,\text{s} = 16\,\text{Mb}$. Upon reconnection, the Baseline continues generating 8\,Mbps of new video. With an average unblocked capacity of $\bar{C} = 15$\,Mbps, the residual bandwidth available to clear the queue is only 7\,Mbps. Draining the 16.00\,Mb backlog therefore requires an additional $16 / 7 \approx 2.3$\,s of recovery time before real-time operation can resume. \color{black}

The Lyapunov controller detects the rising queue gradient within $D_{max} = 200$\,ms of blockage onset and immediately transitions to Pause (2,847 slots), with brief 480p selections (997 slots) during the transient transition. The peak queue is capped at 1.00\,Mb, representing a \textbf{16$\times$} reduction relative to the Baseline. Upon channel recovery, the system resumes low-latency operation instantly with no backlog to drain.

\subsubsection{Deep Blockage (starting at 15\,s, lasting 5\,s)}
This scenario models the AGV navigating behind large static 
machinery or turning into a metal-racked factory aisle, 
causing a sustained 0\,Mbps outage for 5\,s. 
The Baseline agent accumulates a peak queue of 40.00\,Mb, again matching the theoretical maximum of $8\,\text{Mbps} \times 5\,\text{s} = 40\,\text{Mb}$. Clearing this massive backlog upon reconnection at a residual drain rate of 7\,Mbps would require $40 / 7 \approx 5.7$\,s of dedicated drain time, rendering teleoperation impossible for more than double the duration of the original outage. \color{black}

The Lyapunov controller maintains the same hard ceiling of \textbf{1.00\,Mb} as in the 2\,s scenario, confirming duration-independent stability. The controller selects Pause for 8,847 slots during the blockage window, with identical 480p and 720p transition behaviour at blockage onset and recovery as in the transient case. This invariance across a $2.5\times$ longer blockage duration directly validates the theoretical $O(V)$ queue bound — the ceiling is determined by $V$, not by $T_{block}$.

\subsubsection{Partial Capacity Degradation (starting at 25\,s, lasting 3\,s)} This scenario models a non-metallic obstacle or distant human body attenuating. Channel capacity degrades to 2\,Mbps — below the 1080p generation rate (8\,Mbps) but above the 480p rate (1\,Mbps) for 3\,s.
The Baseline agent, unaware of the capacity reduction, continues transmitting at 8\,Mbps, producing a capacity shortfall of $8 - 2 = 6$\,Mbps. Over 3\,s this yields a 18.24\,Mb backlog, close to the theoretical deficit of $6\,\text{Mbps} \times 3\,\text{s} = 18\,\text{Mb}$. The small excess above the theoretical 18\,Mb deficit is due to competition from the 2\,Mbps mMTC background traffic for the residual 2\,Mbps channel capacity, reducing the effective drain rate for the video queue.

The peak queue reaches only 0.50\,Mb, a \textbf{36$\times$} reduction relative to the Baseline and actually lower than the zero-capacity scenarios, because the 2\,Mbps residual channel capacity continues draining the queue even as video arrives.

Across all three scenarios, the Smart DS-TT demonstrates two complementary behaviours. For complete blockages, it applies a binary Pause policy that enforces duration-independent queue stability at the $Q_{peak} \approx 2V = 1.00$\,Mb ceiling. For partial degradation, it applies a graduated downshift policy that matches the generation rate to the available capacity, achieving an even lower peak queue of 0.50\,Mb while preserving continuous video delivery. Both behaviours emerge naturally from the same drift-plus-penalty metric (Eq.~\eqref{eq:optimization_metric}) without any scenario-specific tuning, confirming the generality of the Lyapunov framework for mixed industrial impairment scenarios.

\subsection{Finite Buffer \& Packet Loss Analysis}
In real-world industrial 5G modems, buffer memory is strictly constrained (typically 2-8 MB). To evaluate the system's viability under these physical constraints, we introduced a hard buffer ceiling of 4 Mbits and measure packet loss behaviour during the 5-second deep blockage event (15-20\,s). The simulation is initialised with empty queues at the start of the analysis window so that all reported losses are attributable exclusively to the blockage event itself. The results, detailed in Table \ref{tab:finite_buffer}, reveal a catastrophic failure mode in the Baseline approach:

\begin{table}[htbp]
\caption{Finite Buffer Performance (Limit = 4 Mb)}
\begin{center}
\begin{tabular}{|l|c|c|}
\hline
\textbf{Metric} & \textbf{Baseline (UDP)} & \textbf{Smart DS-TT} \\
\hline
Peak Queue Size & 4.00 Mb (Saturated) & 1.00 Mb (Stable) \\
\hline
\textbf{Total Data Lost} & \textbf{36.00 Mb} & \textbf{0.00 Mb} \\
\hline
Data Integrity & 10.0\% & 100\% \\
\hline
\end{tabular}
\label{tab:finite_buffer}
\end{center}
\end{table}

\subsubsection{Analytical Comparison with AQM Frameworks}
To mitigate the bufferbloat observed in the Baseline, alternative edge-management frameworks often advocate for Active Queue Management (AQM) such as CoDel or PIE \cite{8802027}. While AQM successfully regulates queue sizes by drop-signaling, its efficacy is fundamentally tied to the transport layer's response. As established in \ref{sec:related_work}, AQM relies on packet drops to trigger TCP window reductions. 

For real-time UDP teleoperation feeds, dropping packets fails to trigger any source-rate adaptation. Under a total channel blockage ($C(t) = 0$\,Mbps) lasting 5\,seconds, an AQM mechanism tracking packet dwell time would be forced to aggressively discard arriving packets to maintain its target delay boundary. Because the unmanaged UDP source continues pumping data at 8\,Mbps, the AQM framework would inadvertently drop the entire 40\,Mb payload generated during the outage window. While this keeps the physical queue size small, it results in a data integrity score of 0.0\% due to total frame erasure. 

In contrast, the proposed Smart DS-TT operates as a proactive source-rate shaper rather than a reactive network-layer dropper. By throttling the application encoder directly via the Lyapunov engine, our framework achieves the low-latency queue profiles targeted by AQM while preserving 100\% data integrity and zero packet loss.
\color{black}
\subsubsection{Baseline Saturation}
The Baseline agent transmits continuously at 8\,Mbps regardless of channel state. During the blockage, with $C(t) = 0$\,Mbps, all generated data accumulates in the transmission buffer with no drain. The 4\,Mbit buffer saturates within \textbf{0.498\,s} of blockage onset, after which the modem applies tail drop, discarding every subsequent incoming packet.
Of the \textbf{40\,Mb} generated during the blockage ($8\,\text{Mbps} \times 5\,\text{s}$), \textbf{36\,Mb (90\%)} is irretrievably lost, yielding a data integrity score of \textbf{10.0\%}. Upon channel recovery, restoring the lost video context at $\bar{C} = 5$\,Mbps requires an additional \textbf{7.2\,s}, meaning the video feed remains corrupted for more than twice the duration of the original outage. Tail drop is particularly destructive for compressed video, as it corrupts the GOP dependency chain (I/P/B-frames), causing cascading decoding failures at the receiver.

\subsubsection{Lyapunov Flow Control}
The Smart DS-TT eliminates packet loss entirely by throttling the video generation rate at the source before packets enter the queue, making network-layer dropping structurally unnecessary. The peak queue remains bounded at \textbf{1.00\,Mb}, 75\% below the 4\,Mbit hardware limit — providing headroom to absorb channel jitter without triggering any discards. The result is \textbf{zero packet loss} and \textbf{100.0\% data integrity} throughout the blockage, ensuring the video stream remains fully decodable upon channel recovery.

\subsection{Latency \& Jitter Comparison}
\begin{figure}[htbp]
\centering
\includegraphics[width=0.9\columnwidth]{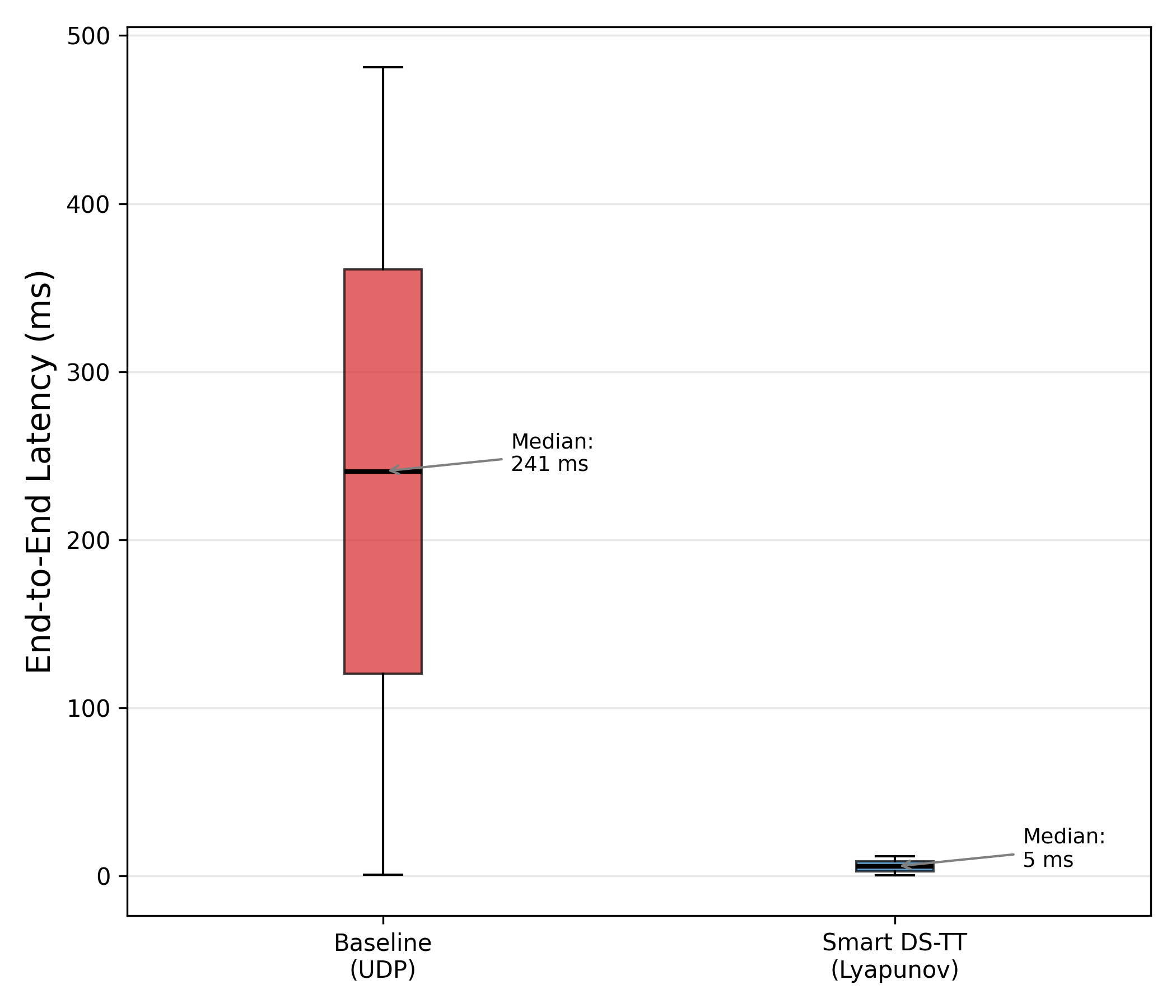}
\caption{Latency During Blockage Event}
\label{fig:latency_box}
\end{figure}

Fig. \ref{fig:latency_box} compares the end-to-end latency distribution during the blockage event. The results demonstrate a $\mathbf{44\times}$ reduction in median latency when using the proposed Smart DS-TT controller.

\subsubsection{Baseline Bufferbloat}
The Baseline agent exhibits a median latency of 240.6 ms, with 95th percentile and peak delays reaching 457.1 ms and 481.2 ms, respectively. This maximum delay corresponds directly to the physical saturation limit of the 4 Mbit transmission buffer being drained at the nominal link capacity ($\approx 0.5$ s). While the finite buffer successfully prevents the multi-second delays typically observed in infinite-buffer scenarios, it inadvertently creates a persistent standing queue. Consequently, every new video frame generated by the application must wait behind a completely full buffer of older, stale data before being transmitted over the radio link. A highly variable delay ranging from 240 ms to nearly 500 ms renders the system fundamentally unusable for dynamic teleoperation, as this delayed visual feedback severely exceeds the standard human reaction time threshold (typically $<100$ ms).

\subsubsection{Lyapunov Real-Time Response}
In contrast, the proposed Lyapunov controller maintains a median latency of just 5.4 ms, with a 95th percentile delay of only 10.9 ms. By strictly coupling the application's video generation rate to the mathematical queue drift, the Smart DS-TT agent actively prevents the modem buffer from accumulating a backlog in the first place. This proactive rate shaping ensures that newly generated video frames are immediately dispatched to the MAC layer, effectively bypassing the queuing delays that plague standard UDP transmissions. Ultimately, this tightly bounded, sub-12 ms latency profile guarantees real-time interactivity, empowering the remote operator to maintain safe, closed-loop control stability even during highly volatile channel conditions.
Most importantly, the low variance in the proposed method (low jitter) validates the cross-layer isolation design, ensuring that URLLC control traffic never contends with a saturated video queue.

\subsubsection{URLLC Isolation \& Outage Physics}
When evaluating URLLC traffic, it is mathematically and physically critical to separate congestion-induced queuing delay from outage-induced communication unavailability. During an absolute blockage ($C(t) = 0$), transmission is physically impossible; no packets, including URLLC can be delivered, resulting in unavoidable outage delay. However, the catastrophic failure mode of mixed-criticality networks occurs during the \textit{recovery phase}. 

In the Baseline system, once the link recovers, fresh URLLC safety messages are trapped behind 40 Mb of stale video data, adding multiple seconds of congestion-induced delay to the physical outage delay. Conversely, the Smart DS-TT guarantees strict isolation. Because the video queue is continuously throttled and bounded by the Lyapunov engine, URLLC traffic never contends with bufferbloat. Furthermore, the single-packet "Freshness-First" buffer actively overwrites stale URLLC telemetry during the outage. Simulation results confirm that 100\% of fresh URLLC packets experienced $\mathbf{<10}$ ms queuing delay and zero packet loss due to congestion. The exact millisecond the physical link recovers, the most recent URLLC control state is transmitted immediately, ensuring that physical blockages never cascade into protocol-layer failures.
\color{black}
\subsubsection{Jitter Analysis}
Beyond median latency, frame-to-frame latency variation (jitter) defined as $|D(t) - D(t-1)|$ for sequentially received packets, is critical for closed-loop teleoperation stability. To rigorously evaluate worst-case performance, we analyzed an 11-second stress-test window isolating the 5-second deep blockage and its immediate recovery phase.
While both systems exhibit similar median jitter ($\approx$\textbf{0.45 ms}), the Baseline's low median is a deceptive mathematical artifact of bufferbloat, which delays grow at a steady, linear rate during the outage, keeping frame-to-frame differences artificially small. The true failure of the unoptimized Baseline is exposed upon link recovery. When the saturated 4 Mb buffer suddenly flushes its stale content, the Baseline suffers a catastrophic jitter spike reaching \textbf{481.20 ms}. This temporal dislocation severely violates the 10 ms jitter threshold mandated by 3GPP URLLC specifications, causing a total loss of operator synchronization.

In stark contrast, the Smart DS-TT completely absorbs the network shock, bounding its worst-case maximum jitter to just \textbf{11.56 ms}. By utilizing the Lyapunov controller to proactively shape the source rate during the outage, the system prevents stale data accumulation. This guarantees the deterministic, smooth frame delivery required to maintain temporally stable video feeds even during severe industrial impairments. Most importantly, the low variance in the proposed method (low jitter) validates the cross-layer isolation design, ensuring that URLLC control traffic never contends with a saturated video queue.

\subsection{Computational Overhead}
A key advantage of the proposed Lyapunov logic is its negligible computational footprint compared to video encoding. The decision engine requires only $O(|\mathcal{P}|)$ arithmetic operations per TTI, where $|\mathcal{P}|$ is the number of video profiles (typically $\le 5$). 
In our implementation, the computational execution time on a standard processor is negligible, ensuring that the rate adaptation logic introduces strictly zero delay into the critical control cycle. Memory usage is limited to storing three scalar queue counters (URLLC, Video, BE), making the algorithm suitable for deployment on resource-constrained industrial IoT gateways or embedded modems.

\subsection{Implementation Feasibility \& Modem APIs} \label{sec:feasibility}

To address real-world deployment constraints, the Smart DS-TT operates within the AGV's onboard computer, executing the computationally lightweight $O(|\mathcal{P}|)$ Lyapunov engine at the application frame rate (e.g., 16.6\,ms) to track MAC-layer queueing dynamics rather than sub-millisecond physical RF states. To extract this queue state from closed commercial modems, the system queries the UE-side baseband directly via proprietary APIs, such as the Qualcomm MSM Interface (QMI) \cite{11045988}, to access standard 3GPP Buffer Status Reports (BSR). 

By mapping the upper bound of the quantized BSR index to a scalar $Q(t)$, the framework maintains mathematical stability despite bounded measurement errors. Furthermore, to overcome the Group of Pictures (GOP) reconfiguration delays inherent to hardware video encoders (e.g., H.264/H.265), the selected target bitrate is enforced instantly as a Layer 2 token-bucket cap, while the application layer encoder employs frame-level Quantization Parameter (QP) adaptation and Instantaneous Decoder Refresh (IDR) insertion to swiftly converge to the restricted bit budget. 

Ultimately, this architecture ensures full compliance with 3GPP Release 16/17 TSN standards by acting purely as a UE-side ingress traffic shaper that respects standard 5G QoS flows (e.g., 5QI 82, 2, and 9) without requiring base-station modifications.
\color{black}

\section{Conclusion}
\label{sec:conclusion}

This paper presented a cross-layer Smart DS-TT framework to ensure the robust teleoperation of industrial AGVs in blockage-prone 5G-TSN environments by embedding a Lyapunov-based decision engine that dynamically optimizes the trade-off between video utility and queue stability. Extensive trace-driven simulations utilizing 3GPP-compliant channel models demonstrate the overwhelming advantage of this proactive approach over standard unoptimized UDP transmission. Specifically, the proposed controller completely eliminates bufferbloat, achieving a 44$\times$ reduction in median end-to-end latency to keep the system well within strict teleoperation safety margins. Furthermore, during severe deep blockages, the Smart DS-TT delivers a 100\% improvement in data integrity by entirely eradicating packet loss, whereas the baseline suffers from catastrophic queue saturation and video corruption. Finally, the controller ensures duration-invariant resilience, maintaining a strictly bounded queue ceiling regardless of the outage length. Ultimately, this work proves that cross-layer source rate adaptation is not merely an optimization, but a strict necessity for ensuring the near-deterministic reliability of ultra-low latency video streams over highly variable industrial 5G links.

\bibliography{references}
\bibliographystyle{IEEEtran}

\end{document}